**The problem of gauge invariance and the functional Schrödinger approach**

by

Dan Solomon

Rauland-Borg Corporation

3450 W. Oakton Street

Skokie, IL 60076

USA

Phone: 1-847-324-8337

Email: dan.Solomon@rauland.com

(Oct 20, 2007)

**Abstract**

Quantum field theory is assumed to be gauge invariant. However it is well known that when certain quantities, such as the vacuum current, are calculated the results are not gauge invariant. The non-gauge invariant terms have to be removed in order to obtain a physically correct result. It has been shown in Ref. [3] and [4] that this problem may be due to a mathematical inconsistency in the canonical formulation of QFT. In this article we will review this previous work and then examine an alternative formulation of QFT called the functional Schrödinger approach. It will be shown that this approach produces different results then the canonical formulation. In particular it will be shown that in the functional Schrödinger approach the vacuum current is gauge invariant.

## 1. Introduction

Quantum field theory is assumed to be gauge invariant [1][2]. However, it is well known that when certain quantities are calculated using perturbation theory the results are not gauge invariant [3].

A change in the gauge is a change in the electromagnetic potential that does not produce a change in the electromagnetic field. The electromagnetic field is given by,

$$\vec{E} = -\left(\frac{\partial \vec{A}}{\partial t} + \vec{\nabla} A_0\right); \quad \vec{B} = \vec{\nabla} \times \vec{A} \tag{1.1}$$

where $\vec{E}$ is the electric field, $\vec{B}$ is the magnetic field, and $\left(A_0, \vec{A}\right)$ is the electromagnetic potential. A change in the electromagnetic potential that does not produce a change the electromagnetic field is given by,

$$\vec{A} \rightarrow \vec{A}' = \vec{A} - \vec{\nabla}\chi, \quad A_0 \rightarrow A_0' = A_0 + \frac{\partial \chi}{\partial t} \tag{1.2}$$

where $\chi\left(\vec{x}, t\right)$ is an arbitrary real valued function. Using relativistic notation this can also be written as,

$$A_\nu \rightarrow A_\nu' = A_\nu + \partial_\nu \chi \tag{1.3}$$

In order for quantum field theory to be gauge invariant a change in the gauge cannot produce a change in any physical observable such as the current and charge expectation values. For example, consider a system that is initially in the vacuum state. The application of an electromagnetic field will perturb this initial state. The current that is induced in the vacuum state due to the application of the electromagnetic field is called the vacuum current $J_{vac}^\mu\left(x\right)$. Using perturbation theory the lowest order term of the vacuum current can be shown to be given by,

$$J_{vac}^\mu\left(x\right) = \int \pi^{\mu\nu}\left(x - x'\right) A_\nu\left(x'\right) d^4x' \tag{1.4}$$

where $\pi^{\mu\nu}$ is called the polarization tensor and summation over repeated indices is assumed. The above relationship is normally written in terms of Fourier transformed quantities as,

$$J_{vac}^\mu\left(k\right) = \pi^{\mu\nu}\left(k\right) A_\nu\left(k\right) \tag{1.5}$$

where k is the 4-momentum. In this case, using relativistic notation, a gauge transformation takes the following form,

$$A_\nu(k) \rightarrow A'_\nu(k) = A_\nu(k) + ik_\nu \chi(k) \tag{1.6}$$

The change in the vacuum current, $\delta_g J^\mu_{vac}(k)$, due to a gauge transformation can be obtained by using (1.6) in (1.5) to yield,

$$\delta_g J^\mu_{vac}(k) = ik_\nu \pi^{\mu\nu}(k)\chi(k) \tag{1.7}$$

Now the vacuum current is an observable quantity therefore, if quantum theory is gauge invariant, the vacuum current must not be affected by a gauge transformation. This means that $\delta_g J^\mu_{vac}(k)$ must be zero. For this to be true we must have that,

$$k_\nu \pi^{\mu\nu}(k) = 0 \tag{1.8}$$

However, when the polarization tensor is calculated it is found that the above relationship does not hold. As discussed in [3] calculations of the polarization tensor by various authors yield,

$$\pi^{\mu\nu}(k) = \pi^{\mu\nu}_G(k) + \pi^{\mu\nu}_{NG}(k) \tag{1.9}$$

In the above expression the quantity $\pi^{\mu\nu}_G(k)$ is gauge invariant, that is, is satisfies $k_\mu \pi^{\mu\nu}_G(k) = 0$. However the quantity $\pi^{\mu\nu}_{NG}(k)$ is not gauge invariant because $k_\mu \pi^{\mu\nu}_G(k) \neq 0$. Therefore to get a physically valid result it is necessary to "correct" equation (1.9) by dropping $\pi^{\mu\nu}_{NG}$ from the solution. This can be done using the process of "regularization" (see Ref [3] and references therein). While this seems to cure the problem by removing the unwanted terms it introduces an additional step that is not in the original formulation of the theory. The obvious question to ask then is why the additional step of regularization is needed in the first place. If the theory is gauge invariant then why does a calculation of the polarization tensor produce non-gauge invariant terms?

In [3] this question was addressed by considering a "simple" field theory consisting of a quantized fermion field in the presence of a classical electromagnetic field. The mathematical consistency of this system was analyzed. Four elements that are normally considered to be part of quantum field theory were examined. These were; (1) that the Schrödinger equation governs the dynamics of the theory with the Hamiltonian specified by Eq. (2.2) of [3]; (2) the theory is gauge invariant; (3) there is local charge conservation, i.e., the continuity equation is true and; (4) there is lower bound to the free field energy. It was shown that these elements of QFT are not mathematically consistent. Specifically item (2) is incompatible with item (4), that is, if QFT is gauge invariant then there cannot be a lower bound to the free field energy. However it can be readily shown that the vacuum state is a lower bound to the free field energy. Therefore, as discussed in [3] and [4], QFT in the Schrödinger picture is not gauge invariant at the formal level. This, then, explains why non-gauge invariant terms appear in the polarization tensor. Since the theory in not gauge invariant in the first place it would be expected that the results of calculations are also not gauge invariant. This is, of course, exactly what happens.

In [3] we considered the problem of gauge invariance in the standard or "canonical" field theory. Here we will consider a different version of field theory called the functional Schrödinger approach. It will be shown that this approach may yield a gauge invariant theory.

The discussion will proceed as follows. In Section 2 we will introduce the basic elements of the canonical field theory in the Schrödinger picture. The key result of this section will be that the free field energy of any state $|\Omega\rangle$ is greater than the free field energy of the vacuum state $|0\rangle$. In Section 3 the gauge invariance of the theory is considered. There is an inconsistency between the key result of Section 2 and the requirements of gauge invariance. The conclusion of this section is that the canonical formulation is not gauge invariant at the formal level. This is, of course, why non-gauge invariant terms appear in the polarization tensor. In Section 4 we examine quantum field theory using the functional Schrödinger approach and verify that when this approach is used a calculation of the vacuum current will be gauge invariant.

## 2. Quantum Field theory.

In this section the basic elements of canonical quantum field theory in the Schrödinger picture will be introduced. We shall consider a "simple" field theory consisting of non-interacting fermions acted on by a classical electromagnetic field.

Natural units will be used so that $\hbar = c = 1$. The time evolution of the state vector $\left| \Omega(t) \right\rangle$ and its dual $\left\langle \Omega(t) \right|$ are given by,

$$i \frac{\partial \left| \Omega(t) \right\rangle}{\partial t} = \hat{H} \left| \Omega(t) \right\rangle, \quad -i \frac{\partial \left\langle \Omega(t) \right|}{\partial t} = \left\langle \Omega(t) \right| \hat{H} \tag{2.1}$$

where $\hat{H}$ is the Hamiltonian operator which is given by,

$$\hat{H} = \int \psi^\dagger(\vec{x}) H \psi(\vec{x}) d\vec{x} - \varepsilon_r \tag{2.2}$$

where $\varepsilon_r$ is a renormalization constant so the energy of the vacuum state is zero and the field operators $\psi(\vec{x})$ obey the equal time anticommutation relationships,

$$\psi_a^\dagger(\vec{x}) \psi_b(\vec{y}) + \psi_b(\vec{y}) \psi_a^\dagger(\vec{x}) = \delta_{ab} \delta^{(3)}(\vec{x} - \vec{y}) \tag{2.3}$$

with all other anticommutators being equal to zero. In the above expression $a$ and $b$ are spinnor indices and,

$$H(A_0, \vec{A}) = H_0 - q \vec{\alpha} \cdot \vec{A} + q A_0 \tag{2.4}$$

with,

$$H_0 = -i \vec{\alpha} \cdot \vec{\nabla} + \beta m \tag{2.5}$$

In the above expression the electromagnetic potential $\left( A_0(\vec{x}, t), \vec{A}(\vec{x}, t) \right)$ is taken to be a classical, unquantized, real valued quantity. Also $q$ and $m$ are the charge and mass of the electron, respectively, and $\vec{\alpha}$ and $\beta$ are the usual 4x4 matrices.

We can also write (2.2) as,

$$\hat{H} = \hat{H}_0 - \int \hat{\vec{J}} \cdot \vec{A} d\vec{x} + \int \hat{\rho} A_0 d\vec{x} \tag{2.6}$$

where,

$$\hat{H}_0 = \int \hat{\psi}_0^\dagger(\vec{x}) H_0 \hat{\psi}_0(\vec{x}) d\vec{x} - \varepsilon_r \tag{2.7}$$

and charge and current operators are defined by,

$$\hat{\rho}(\vec{x}) = q \psi^\dagger(\vec{x}) \psi(\vec{x}) \text{ and } \hat{\vec{J}}(\vec{x}) = q \psi^\dagger(\vec{x}) \vec{\alpha} \psi(\vec{x}) \tag{2.8}$$

Let $\left| n \right\rangle$ designate the eigenstates of the operator $\hat{H}_0$ with eigenvalues $\varepsilon_n$. They from a complete orthonormal basis and satisfy the relationships,

$$\hat{H}_o \left| n \right\rangle = \varepsilon_n \left| n \right\rangle; \quad \left\langle n \right| \hat{H}_o = \left\langle n \right| \varepsilon_n \tag{2.9}$$

and,

$$\left\langle n \middle| m \right\rangle = \delta_{nm} \tag{2.10}$$

and,

$$\sum_n \left| n \right\rangle \left\langle n \right| = 1 \tag{2.11}$$

The vacuum state $\left| 0 \right\rangle$ is assumed to be the eigenvector of $\hat{H}_0$ with the smallest eigenvalue $\varepsilon_o = 0$. For all other eigenvalues,

$$\varepsilon_n > \varepsilon_o = 0 \text{ for } \left| n \right\rangle \neq \left| 0 \right\rangle \tag{2.12}$$

Any arbitrary normalized state $\left| \Omega \right\rangle$ can be expanded in terms of the eigenstates $\left| n \right\rangle$ so that we can write,

$$\left| \Omega \right\rangle = \sum_n c_n \left| n \right\rangle \tag{2.13}$$

where $c_n$ are the expansion coefficients and where the normalization condition is,

$$\left\langle \Omega \middle| \Omega \right\rangle = \sum_n \left| c_n \right|^2 = 1 \tag{2.14}$$

The expression $\left\langle \Omega \middle| \hat{H}_0 \middle| \Omega \right\rangle$ is called the free field energy of the normalized state $\left| \Omega \right\rangle$ and is simply the energy when the electromagnetic potential is zero. Using the above expressions we can show that,

$$\left\langle \Omega \middle| \hat{H}_0 \middle| \Omega \right\rangle = \sum_n \left| c_n \right|^2 \varepsilon_n \tag{2.15}$$

Using this fact along with (2.12) we can easily show that,

$$\left\langle \Omega \middle| \hat{H}_0 \middle| \Omega \right\rangle > \left\langle 0 \middle| \hat{H}_0 \middle| 0 \right\rangle = 0 \text{ for all } \left| \Omega \right\rangle \neq \left| 0 \right\rangle \tag{2.16}$$

Therefore the vacuum state is the quantum state with the minimum value of the free field energy.

**The Schrödinger picture and gauge invariance.**

Now we examine the question of whether or not the canonical approach is gauge invariant. If has already been shown in Ref. [3] that it is not gauge invariant. Here we will present an alternative proof of this fact. The proof is similar to that of Ref. [4].

If the theory is gauge invariant then a change in the gauge should not change the value of the physical observables. These include the charge and current expectation values which are defined by,

$$\rho_e\left(\vec{x},t\right) = \left\langle \Omega(t) \middle| \hat{\rho}\left(\vec{x}\right) \middle| \Omega(t) \right\rangle \text{ and } \vec{J}_e\left(\vec{x},t\right) = \left\langle \Omega(t) \middle| \hat{\vec{J}}\left(\vec{x}\right) \middle| \Omega(t) \right\rangle \tag{3.1}$$

respectively,

To test this assume that at the initial time $t=0$ the system is in the vacuum state $\left|0\right\rangle$ and the electromagnetic potential is zero. Now let us consider two problems. In the first let the state vector involve forward in time in the presence of the electromagnetic potential,

$$\left( A_0^{(1)}\left(\vec{x},t\right), \vec{A}^{(1)}\left(\vec{x},t\right) \right) = 0 \tag{3.2}$$

In this case the state vector $\left|\Omega_1\left(t\right)\right\rangle$ obeys the equation,

$$i\frac{\partial \left|\Omega_1\left(t\right)\right\rangle}{\partial t} = \hat{H}_0 \left|\Omega_1\left(t\right)\right\rangle \tag{3.3}$$

and satisfies the initial condition $\left|\Omega_1\left(0\right)\right\rangle = \left|0\right\rangle$. Based on the result of the last section the solution to the above is,

$$\left|\Omega_1\left(t\right)\right\rangle = \left|0\right\rangle \tag{3.4}$$

Therefore the current and charge expectation values are,

$$\vec{J}_{1,e}\left(\vec{x},t\right) = \left\langle 0 \middle| \hat{\vec{J}}\left(\vec{x}\right) \middle| 0 \right\rangle \qquad \rho_{1,e}\left(\vec{x},t\right) = \left\langle 0 \middle| \hat{\rho}\left(\vec{x}\right) \middle| 0 \right\rangle \tag{3.5}$$

Now, for the second problem, let the system evolve forward in time in the presence of an electromagnetic potential given by,

$$\left( A_0^{(2)}\left(\vec{x},t\right), \vec{A}^{(2)}\left(\vec{x},t\right) \right) = \left( \frac{\partial \chi\left(\vec{x},t\right)}{\partial t}, -\vec{\nabla}\chi\left(\vec{x},t\right) \right) \tag{3.6}$$

where $\chi\left(\vec{x},t\right)$ is an arbitrary function that satisfies the initial condition, at $t=0$,

$$\chi\left(\vec{x},0\right)=0 \ \text{ and } \ \frac{\partial\chi\left(\vec{x},t\right)}{\partial t}\bigg|_{t=0}=0 \tag{3.7}$$

Use this in (2.1) along with (2.6) to show that for this problem the state vector $\left|\Omega_2\left(t\right)\right\rangle$

obeys the equation,

$$i\frac{\partial\left|\Omega_2\left(t\right)\right\rangle}{\partial t}=\left(\hat{H}_0+\int\hat{\vec{J}}\cdot\vec{\nabla}\chi d\vec{x}+\int\hat{\rho}\frac{\partial\chi}{\partial t}d\vec{x}\right)\left|\Omega_2\left(t\right)\right\rangle \tag{3.8}$$

with the initial condition $\left|\Omega_2\left(0\right)\right\rangle=\left|0\right\rangle$.

Using (3.6) in (2.6) we write the Schrödinger picture Hamiltonian operator as,

$$\hat{H}\left(t\right)=\hat{H}_0+\int\hat{\vec{J}}\left(\vec{x}\right)\cdot\nabla\chi\left(\vec{x},t\right)d\vec{x}+\int\hat{\rho}\left(\vec{x}\right)\frac{\partial\chi\left(\vec{x},t\right)}{\partial t}d\vec{x} \tag{3.9}$$

Next consider the quantity $\left\langle\Omega_2\left(t\right)\middle|\hat{H}_0\middle|\Omega_2\left(t\right)\right\rangle$. Using (3.9) we can obtain the expression,

$$\left\langle\Omega_2\left(t\right)\middle|\hat{H}_0\middle|\Omega_2\left(t\right)\right\rangle=\left\langle\Omega_2\left(t\right)\middle|\left(\hat{H}\left(t\right)-\int\hat{J}\left(\vec{x}\right)\cdot\nabla\chi\left(\vec{x},t\right)d\vec{x}-\int\hat{\rho}\left(\vec{x}\right)\frac{\partial\chi\left(\vec{x},t\right)}{\partial t}d\vec{x}\right)\middle|\Omega_2\left(t\right)\right\rangle \tag{3.10}$$

This can also be written as,

$$\left\langle\Omega_2\left(t\right)\middle|\hat{H}_0\middle|\Omega_2\left(t\right)\right\rangle=\left\langle\Omega_2\left(t\right)\middle|\hat{H}\left(t\right)\middle|\Omega_2\left(t\right)\right\rangle-\int\vec{J}_{2,e}\left(\vec{x},t\right)\cdot\nabla\chi\left(\vec{x},t\right)d\vec{x}-\int\rho_{2,e}\left(\vec{x},t\right)\frac{\partial\chi\left(\vec{x},t\right)}{\partial t}d\vec{x} \tag{3.11}$$

where $\vec{J}_{2,e}\left(\vec{x},t\right)$ and $\rho_{2,e}\left(\vec{x},t\right)$ are the current and charge expectation values,

respectively, and are given by,

$$\vec{J}_{2,e}\left(\vec{x},t\right)=\left\langle\Omega_2\left(t\right)\middle|\hat{\vec{J}}\left(\vec{x}\right)\middle|\Omega_2\left(t\right)\right\rangle \ \text{ and } \ \rho_{2,e}\left(\vec{x},t\right)=\left\langle\Omega_2\left(t\right)\middle|\hat{\rho}\left(\vec{x}\right)\middle|\Omega_2\left(t\right)\right\rangle \tag{3.12}$$

Next use (2.1) to obtain,

$$\frac{\partial}{\partial t}\left\langle\Omega_2\left(t\right)\middle|\hat{H}\left(t\right)\middle|\Omega_2\left(t\right)\right\rangle=\left\langle\Omega_2\left(t\right)\middle|\frac{\partial\hat{H}\left(t\right)}{\partial t}\middle|\Omega_2\left(t\right)\right\rangle \tag{3.13}$$

Take the time derivative of (3.11) and use the above expression along with,

$$\frac{\partial}{\partial t}\hat{H}\left(t\right)=\int\hat{\vec{J}}\left(\vec{x}\right)\cdot\nabla\frac{\partial\chi\left(\vec{x},t\right)}{\partial t}d\vec{x}+\int\hat{\rho}\left(\vec{x}\right)\frac{\partial^2\chi\left(\vec{x},t\right)}{\partial t^2}d\vec{x} \tag{3.14}$$

to obtain,

$$\frac{\partial \langle \Omega_2(t) | \hat{H}_0 | \Omega_2(t) \rangle}{\partial t} = -\int \frac{\partial \vec{J}_{2,e}(\vec{x},t)}{\partial t} \cdot \nabla \chi(\vec{x},t) \, d\vec{x} - \int \frac{\partial \hat{\rho}_{2,e}(\vec{x},t)}{\partial t} \frac{\partial \chi(\vec{x},t)}{\partial t} \, d\vec{x} \qquad (3.15)$$

Now we will invoke the principle of gauge invariance. Refer to (1.2) to show that the electromagnetic potential $\left( A_0^{(1)}(\vec{x},t), \vec{A}^{(1)}(\vec{x},t) \right)$ and $\left( A_0^{(2)}(\vec{x},t), \vec{A}^{(2)}(\vec{x},t) \right)$ are related by a gauge transformation. Therefore, if the theory is gauge invariant, the current and charge expectation values must by the same in both cases. Use this fact along with (3.5) to obtain,

$$\vec{J}_{2,e}(\vec{x},t) = \vec{J}_{1,e}(\vec{x},t) = \langle 0 | \hat{\vec{J}}(\vec{x}) | 0 \rangle \ \text{ and } \ \rho_{2,e}(\vec{x},t) = \rho_{1,e}(\vec{x},t) = \langle 0 | \hat{\rho}(\vec{x}) | 0 \rangle \quad (3.16)$$

Now it is evident from the above expressions that $\vec{J}_{2,e}(\vec{x},t)$ and $\rho_{2,e}(\vec{x},t)$ are time independent. Use this fact in (3.15) to obtain,

$$\frac{\partial \langle \Omega_2(t) | \hat{H}_0 | \Omega_2(t) \rangle}{\partial t} = 0 \qquad (3.17)$$

This yields,

$$\langle \Omega_2(t) | \hat{H}_0 | \Omega_2(t) \rangle = \langle \Omega_2(0) | \hat{H}_0 | \Omega_2(0) \rangle \qquad (3.18)$$

Next use the initial condition $| \Omega_2(0) \rangle = | 0 \rangle$ to obtain,

$$\langle \Omega_2(t) | \hat{H}_0 | \Omega_2(t) \rangle = \langle 0 | \hat{H}_0 | 0 \rangle \qquad (3.19)$$

Now referring to Eqs. (3.3) and (3.8) it is evident that $| \Omega_2(t) \rangle \neq | \Omega_1(t) \rangle = | 0 \rangle$ because they solve different differential equations. Therefore Eq. (3.19) is in direct contradiction to (2.16). According to (2.16) the free field energy of state $| \Omega_2(t) \rangle$ should be greater than the free field energy of the vacuum state $| 0 \rangle$ since $| \Omega_2(t) \rangle \neq | 0 \rangle$. However according to (3.19) the free field energies are equal. Eq. (3.19) was obtained by assuming that the theory was gauge invariant. Since this assumption leads to a contradiction then the canonical form of the Schrödinger picture cannot be gauge invariant.

This result is consistent with the results of Ref. [3] and [4]. It also explains why non-gauge invariant terms appear in the polarization tensor prior to the step of regularization. Since the formal theory is not gauge invariant then calculations derived from the formal theory should not be expected to be gauge invariant either.

## 4. Functional Schrödinger equation

We have shown that the canonical formulation of QFT does not yield a gauge invariant theory. However there another way to approach QFT. This alternative approach is based on the functional Schrödinger equation. A detailed derivation of the functional Schrödinger equation is given in references [5], [6], and [7]. Some of the basic elements will be outlined here. First replace the field operators by the following quantities,

$$\hat{\psi}_\alpha(\vec{x}) \rightarrow \frac{1}{\sqrt{2}}\left(u_\alpha(\vec{x}) + \frac{\delta}{\delta u_\alpha^\dagger(\vec{x})}\right); \quad \hat{\psi}_\alpha^\dagger(\vec{x}) \rightarrow \frac{1}{\sqrt{2}}\left(u_\alpha^\dagger(\vec{x}) + \frac{\delta}{\delta u_\alpha(\vec{x})}\right) \tag{4.1}$$

where the u and $u^\dagger$ are Grassman variables. When this is done the anti-commutation relationships (2.3) are still valid. Next replace the state vector $\left|\Omega(t)\right\rangle$ by $\Psi\left(u, u^\dagger, t\right)$ which is a function of the Grassman variables. The original Schrödinger equation (Eq. (2.1)) can then be replaced by the functional Schrödinger equation,

$$i\frac{\partial \Psi\left(u, u^\dagger, t\right)}{\partial t} = H_f \Psi\left(u, u^\dagger, t\right) \tag{4.2}$$

where,

$$H_f = \frac{1}{2}\int d\vec{x}d\vec{y}\left[\left(u^\dagger(\vec{x}) + \frac{\delta}{\delta u(\vec{x})}\right)h(\vec{x},\vec{y})\left(u(\vec{y}) + \frac{\delta}{\delta u^\dagger(\vec{y})}\right)\right] \tag{4.3}$$

and where,

$$h(\vec{x},\vec{y}) = h_0(\vec{x},\vec{y}) - \left(q\vec{\alpha}\cdot\vec{A} - qA_0\right)\delta^3(\vec{x}-\vec{y}) \tag{4.4}$$

and

$$h_0(\vec{x},\vec{y}) = -i\vec{\alpha}\cdot\vec{\nabla}_x\delta^3(\vec{x}-\vec{y}) + \beta m\delta^3(\vec{x}-\vec{y}) = H_0(\vec{x})\delta^3(\vec{x}-\vec{y}) \tag{4.5}$$

Note that the subscript "f" on the term $H_f$ is just a reminder that this is the Hamiltonian in the functional Schrödinger representation.

It is shown by [5] that the vacuum state is given by,

$$\Psi_v = N\exp\left(\int d\vec{x}d\vec{y}u^\dagger(\vec{x})G_v(\vec{x},\vec{y})u(\vec{y})\right) \tag{4.6}$$

where N a is constant and,

$$G_\nu\left(\vec{x},\vec{y}\right) = P_-\left(\vec{x},\vec{y}\right) - P_+\left(\vec{x},\vec{y}\right) \tag{4.7}$$

where,

$$P_-\left(\vec{x},\vec{y}\right) = \sum_{\vec{p},s} \varphi^{(0)}_{-1,\vec{p},s}\left(\vec{x}\right)\varphi^{(0)\dagger}_{-1,\vec{p},s}\left(\vec{y}\right) \text{ and } P_+\left(\vec{x},\vec{y}\right) = \sum_{\vec{p},s} \varphi^{(0)}_{+1,\vec{p},s}\left(\vec{x}\right)\varphi^{(0)\dagger}_{+1,\vec{p},s}\left(\vec{y}\right) \tag{4.8}$$

(Note that Ref. [5] uses the symbol '$\Omega$' instead of 'G'). In the above expression the $\varphi^{(0)}_{\lambda,\vec{p},s}\left(\vec{x}\right)$ are the eigensolutions of the operator $H_0$ and satisfy,

$$H_0\varphi^{(0)}_{\lambda,\vec{p},s}\left(\vec{x}\right) = \lambda E_{\vec{p}}\varphi^{(0)}_{\lambda,\vec{p},s}\left(\vec{x}\right) \tag{4.9}$$

where,

$$E_{\vec{p}} = +\sqrt{\vec{p}^2+m^2}, \quad \lambda = \begin{vmatrix} +1 \text{ for a positive energy state} \\ -1 \text{ for a negative energy state} \end{vmatrix} \tag{4.10}$$

and where $\vec{p}$ is the momentum of the state and $s = \pm 1/2$ is the spin index.

The $\varphi^{(0)}_{\lambda,\vec{p},s}\left(\vec{x}\right)$ can be expressed by,

$$\varphi^{(0)}_{\lambda,\vec{p},s}\left(\vec{x}\right) = u_{\lambda,\vec{p},s}e^{i\vec{p}\cdot\vec{x}} \tag{4.11}$$

where $u_{\lambda,\vec{p},s}$ is a constant 4-spinor which are given in Chapt. 2 of Ref. [2]. The $\varphi^{(0)}_{\lambda,\vec{p},s}\left(\vec{x}\right)$ form a complete orthonormal basis in Hilbert space and satisfy

$$\int \varphi^{(0)\dagger}_{\lambda,\vec{p},s}\left(\vec{x}\right)\varphi^{(0)}_{\lambda',\vec{p}',s'}\left(\vec{x}\right)d\vec{x} = \delta_{\lambda,\lambda'}\delta_{s,s'}\delta_{\vec{p},\vec{p}'} \tag{4.12}$$

We can use the above to obtain the following relationships,

$$P_-\left(\vec{x},\vec{y}\right) = \int P_-\left(\vec{x},\vec{z}\right)P_-\left(\vec{z},\vec{y}\right)d\vec{z}; \quad P_+\left(\vec{x},\vec{y}\right) = \int P_+\left(\vec{x},\vec{z}\right)P_+\left(\vec{z},\vec{y}\right)d\vec{z} \tag{4.13}$$

and,

$$\int P_-\left(\vec{x},\vec{z}\right)P_+\left(\vec{z},\vec{y}\right)d\vec{z} = \int P_+\left(\vec{x},\vec{z}\right)P_-\left(\vec{z},\vec{y}\right)d\vec{z} = 0 \tag{4.14}$$

We can consider the quantities $G_\nu$, $P_+$, and $P_-$ to be square matrices and the u and $u^\dagger$ as column matrices and row matrices, respectively. In this case we can write the above relationships using an abbreviated notation as follows,

$$\int d\vec{x}d\vec{y}u^\dagger\left(\vec{x}\right)G_\nu\left(\vec{x},\vec{y}\right)u\left(\vec{y}\right) = u^\dagger G_\nu u \tag{4.15}$$

and,

$$G_\nu = P_- - P_+; \quad P_- = \left(P_-\right)^2; \quad P_+ = \left(P_+\right)^2; \quad P_-P_+ = P_+P_- = 0 \tag{4.16}$$

Also,

$$\left(G_v\right)^2 = \left(P_-\right)^2 + \left(P_+\right)^2 = I \tag{4.17}$$

where I is the unit square matrix.

Now consider a system that is initially in the vacuum state $\Psi_v$ with the electric potential equal to zero. At the some time $t_0$ apply an electromagnetic potential $\left(A_0\left(\vec{x},t\right),\vec{A}\left(\vec{x},t\right)\right)$. The wave functional will evolve forward in time according to Eq. (4.2). At time $t \geq t_0$ the state will be of the form $\Psi_0\left(t\right)$ where,

$$\Psi_0\left(t\right) = N\left(t\right)\exp\left(\int d\vec{x}d\vec{y}u^\dagger\left(\vec{x}\right)G\left(\vec{x},\vec{y},t\right)u\left(\vec{y}\right)\right) \tag{4.18}$$

where the quantity $G\left(t\right)$ evolves according to the equation,

$$i\frac{\partial G\left(t\right)}{\partial t} = \frac{1}{2}\left(I - G\left(t\right)\right)h\left(I + G\left(t\right)\right) \tag{4.19}$$

with the initial condition $G\left(t_0\right) = G_v$ (see Ref. [5]).

The current expectation value in the functional Schrödinger representation for the state $\Psi_0\left(t\right)$ is,

$$\vec{J}_{fe}\left(\vec{x},t\right) = \frac{\left\langle \Psi_0\left(t\right)\middle| \hat{\vec{J}}_f\left(\vec{x}\right)\middle|\Psi_0\left(t\right)\right\rangle}{\left\langle \Psi_0\left(t\right)\middle|\Psi_0\left(t\right)\right\rangle} \tag{4.20}$$

where the current operator in the functional Schrödinger approach is given by,

$$\hat{\vec{J}}_f\left(\vec{x}\right) = \frac{1}{2}\left(u^\dagger\left(\vec{x}\right) + \frac{\delta}{\delta u\left(\vec{x}\right)}\right)\vec{\alpha}\left(u\left(\vec{x}\right) + \frac{\delta}{\delta u^\dagger\left(\vec{x}\right)}\right) \tag{4.21}$$

This can also be written as,

$$\hat{\vec{J}}_f\left(\vec{x}\right) = \frac{1}{2}\int d\vec{y}\left[\left(u^\dagger\left(\vec{x}\right) + \frac{\delta}{\delta u\left(\vec{x}\right)}\right)\vec{\alpha}\delta^{(3)}\left(\vec{x} - \vec{y}\right)\left(u\left(\vec{y}\right) + \frac{\delta}{\delta u^\dagger\left(\vec{y}\right)}\right)\right] \tag{4.22}$$

Referring back to (4.20) we define,

$$\left\langle \Psi_0\middle|\hat{\vec{J}}_f\left(\vec{x}\right)\middle|\Psi_0\right\rangle = \int Du^\dagger Du\Psi_0^\dagger\hat{\vec{J}}_f\left(\vec{x}\right)\Psi_0; \;\; \left\langle\Psi_0\middle|\Psi_0\right\rangle = \int Du^\dagger Du\Psi_0^\dagger\Psi_0 \tag{4.23}$$

where,

$$Du^\dagger Du = \prod_j du\left(\vec{x}_j\right)^\dagger du\left(\vec{x}_j\right) \tag{4.24}$$

Further details are provided in Ref. 5 and 6.

From the discussion in [6] it is shown that when (4.22) is used in (4.20) we obtain,

$$\vec{J}_{fe}(\vec{x},t) = \vec{\alpha}_{AB} \frac{\langle \Psi_0(t) | \psi_A^\dagger \psi_B | \Psi_0(t) \rangle}{\langle \Psi_0(t) | \Psi_0(t) \rangle} = \frac{1}{2} \vec{\alpha}_{AB} R_{BA}(t) \qquad (4.25)$$

where,

$$\vec{\alpha}_{AB} = \vec{\alpha} \delta^3 (\vec{x}_A - \vec{x}_B); \ \hat{\psi}_B = \frac{1}{\sqrt{2}} \left( u_B + \frac{\delta}{\delta u_B^\dagger} \right); \ \hat{\psi}_A^\dagger = \frac{1}{\sqrt{2}} \left( u_A^\dagger + \frac{\delta}{\delta u_A} \right) \qquad (4.26)$$

and $R_{BA}(t)$ is the called the two point function and is given by,

$$R(t) = (I + G(t))(G(t) + \overline{G}(t))^{-1}(I + \overline{G}(t)) \qquad (4.27)$$

where $\overline{G}(t) \equiv \left( G(t)^\dagger \right)^{-1}$.

Therefore in order to determine $\vec{J}_{fe}(\vec{x},t)$ we must solve (4.19) for $G(t)$. From Ref [5] the solution to (4.19) is,

$$G(t) = (Q(t) - P_+)(Q(t) + P_+)^{-1} = 2Q(t)(Q(t) + P_+)^{-1} - I \qquad (4.28)$$

where,

$$i \frac{\partial Q(t)}{\partial t} = hQ(t) \qquad (4.29)$$

and $Q(t)$ satisfies the initial condition $Q(t_0) = P_-$. Therefore $Q(t)$ is given by,

$$Q(\vec{x}, \vec{y}, t) = \sum_{\vec{p},s} \varphi_{-1,\vec{p},s}(\vec{x},t) \varphi_{-1,\vec{p},s}^{(0)\dagger}(\vec{y}) \qquad (4.30)$$

where,

$$i \frac{\partial \varphi_{\lambda,\vec{p},s}(\vec{x},t)}{\partial t} = H(A_0, \vec{A}) \varphi_{\lambda,\vec{p},s}(\vec{x},t) \qquad (4.31)$$

with the initial condition $\varphi_{\lambda,\vec{p},s}(\vec{x},t_0) = \varphi_{\lambda,\vec{p},s}^{(0)}(\vec{x})$. Note that,

$$Q^\dagger(t)Q(t) = P_-; \ \ Q(t)P_+ = 0; \ \ Q(t)P_- = Q(t) \qquad (4.32)$$

where we have used (4.30) and the fact that equations (4.12) and (4.31) ensure that the $\varphi_{\lambda,\vec{p},s}(\vec{x},t)$ form an orthonormal set, that is,

$$\int \varphi^{\dagger}_{\lambda, \vec{p}, s}\left(\vec{x}, t\right) \varphi_{\lambda', \vec{p}', s'}\left(\vec{x}, t\right) d\vec{x} = \delta_{\lambda \lambda'} \delta_{\vec{p} \vec{p}'} \delta_{ss'}$$ (4.33)

Using (4.19) and the initial condition $G\left(t_0\right) = G_v$ we show in the Appendix that

$G\left(t\right)^2 = G_v^2 = I$. From this we obtain $G\left(t\right) = G\left(t\right)^{-1}$ and $G\left(t\right)^{\dagger} = \left(G\left(t\right)^{\dagger}\right)^{-1}$ so that

$\bar{G}\left(t\right) = \left(G\left(t\right)^{\dagger}\right)^{-1} = G\left(t\right)^{\dagger}$. Therefore,

$$R\left(t\right) = \left(I + G\left(t\right)\right)\left(G\left(t\right) + G\left(t\right)^{\dagger}\right)^{-1}\left(I + G\left(t\right)^{\dagger}\right)$$ (4.34)

Use (4.28) in the above to obtain,

$$R\left(t\right) = 4Q\left(t\right)\left(Q\left(t\right) + P_+\right)^{-1}\left(\begin{array}{l}\left(Q\left(t\right) - P_+\right)\left(Q\left(t\right) + P_+\right)^{-1} \\ + \left(Q^{\dagger}\left(t\right) + P_+^{\dagger}\right)^{-1}\left(Q^{\dagger}\left(t\right) - P_+^{\dagger}\right)\end{array}\right)^{-1}\left(Q^{\dagger}\left(t\right) + P_+^{\dagger}\right)^{-1}Q^{\dagger}\left(t\right)$$ (4.35)

Use the matrix relationship $C^{-1}B^{-1}A^{-1} = \left(ABC\right)^{-1}$ to obtain,

$$R\left(t\right) = 4Q\left(t\right)\left(\begin{array}{l}\left(Q^{\dagger}\left(t\right) + P_+^{\dagger}\right)\left(Q\left(t\right) - P_+\right) \\ + \left(Q^{\dagger}\left(t\right) - P_+^{\dagger}\right)\left(Q\left(t\right) + P_+\right)\end{array}\right)^{-1}Q^{\dagger}\left(t\right)$$ (4.36)

This becomes,

$$R\left(t\right) = 4Q\left(t\right)\left(2Q^{\dagger}\left(t\right)Q\left(t\right) - 2P_+^{\dagger}P_+\right)^{-1}Q^{\dagger}\left(t\right) = 2Q\left(t\right)\left(Q^{\dagger}\left(t\right)Q\left(t\right) - P_+\right)^{-1}Q^{\dagger}\left(t\right)$$ (4.37)

where we have used $P_+^{\dagger}P_+ = P_+$. Use (4.32) in the above to obtain,

$$R\left(t\right) = 2Q\left(t\right)\left(P_- - P_+\right)^{-1}Q^{\dagger}\left(t\right) = 2Q\left(t\right)\left(P_- - P_+\right)Q^{\dagger}\left(t\right) = 2Q\left(t\right)Q^{\dagger}\left(t\right)$$ (4.38)

where we have used $\left(P_- - P_+\right)^{-1} = \left(P_- - P_+\right)$. Use (4.30) along with (4.12) in the above to yield,

$$\begin{aligned}\frac{1}{2}R\left(\vec{x}, \vec{y}, t\right) &= \int\left(\sum_{\vec{p}, s}\varphi_{-1, \vec{p}, s}\left(\vec{x}, t\right)\varphi^{(0)\dagger}_{-1, \vec{p}, s}\left(\vec{z}\right)\sum_{\vec{p}', s'}\varphi^{(0)}_{-1, \vec{p}', s'}\left(\vec{z}\right)\varphi^{\dagger}_{-1, \vec{p}', s'}\left(\vec{y}, t\right)\right)d\vec{z} \\ &= \sum_{\vec{p}, s}\varphi_{-1, \vec{p}, s}\left(\vec{x}, t\right)\varphi^{\dagger}_{-1, \vec{p}, s}\left(\vec{y}, t\right)\end{aligned}$$ (4.39)

Use this in (4.25) to yield,

$$\vec{J}_{fe}(\vec{x},t) = \int d\vec{y} \left( \sum_{\vec{p},s} \varphi_{-1,\vec{p},s}(\vec{x},t) \vec{\alpha} \delta^{(3)}(\vec{x}-\vec{y}) \varphi_{-1,\vec{p},s}^{\dagger}(\vec{y},t) \right) = \sum_{\vec{p},s} \varphi_{-1,\vec{p},s}^{\dagger}(\vec{x},t) \vec{\alpha} \varphi_{-1,\vec{p},s}(\vec{x},t)$$

(4.40)

Now we have an expression for the current. Recall that $\vec{J}_{fe}(\vec{x},t)$ is the vacuum current because it is the current induced in a state that was initially in the vacuum state by an applied electromagnetic field. As we have already discussed a calculation of the vacuum current using the canonical formulation contains non-gauge invariant terms. The question we want to address now is whether $\vec{J}_{fe}(\vec{x},t)$ is gauge invariant. To determine this calculate $\vec{J}_{fe}(\vec{x},t)$ for the gauge transformed electromagnetic potential given by (1.2) where $\chi(\vec{x},t)$ satisfies the initial conditions at $t=t_0$,

$$\chi(\vec{x},t_0) = 0 \text{ and } \frac{\partial \chi(\vec{x},t_0)}{\partial t_0} = 0$$

(4.41)

Use the same analysis as that resulting in (4.40) to obtain,

$$\vec{J}_{g,fe}(\vec{x},t) = \sum_{\vec{p},s} \varphi_{-1,\vec{p},s}^{(g)\dagger}(\vec{x},t) \vec{\alpha} \varphi_{-1,\vec{p},s}^{(g)}(\vec{x},t)$$

(4.42)

where $\vec{J}_{g,fe}(\vec{x},t)$ is the gauge transformed current and where $\varphi_{-1,\vec{p},s}^{(g)}(\vec{x},t)$ satisfies the initial condition $\varphi_{-1,\vec{p},s}^{(g)}(\vec{x},0) = \varphi_{-1,\vec{p},s}^{(0)}(\vec{x},0)$ and the Schrödinger equation,

$$i \frac{\partial \varphi_{-1,\vec{p},s}^{(g)}(\vec{x},t)}{\partial t} = H\left(A_0 + \partial \chi/\partial t, \vec{A} - \vec{\nabla}\chi\right) \varphi_{-1,\vec{p},s}^{(g)}(\vec{x},t)$$

(4.43)

It can be readily shown that the solution to this equation is,

$$\varphi_{-1,\vec{p},s}^{(g)}(\vec{x},t) = e^{-iq\chi} \varphi_{-1,\vec{p},s}(\vec{x},t)$$

(4.44)

Recall that $\varphi_{-1,\vec{p},s}(\vec{x},t)$ satisfies (4.31). Use this result in (4.42) to obtain,

$$\vec{J}_{g,fe}(\vec{x},t) = \sum_{\vec{p},s} \varphi_{-1,\vec{p},s}^{\dagger}(\vec{x},t) e^{iq\chi} \vec{\alpha} e^{-iq\chi} \varphi_{-1,\vec{p},s}(\vec{x},t) = \sum_{\vec{p},s} \varphi_{-1,\vec{p},s}^{\dagger}(\vec{x},t) \vec{\alpha} \varphi_{-1,\vec{p},s}(\vec{x},t)$$

(4.45)

Compare this to (4.40) to show that $\vec{J}_{g,fe}(\vec{x},t) = \vec{J}_{fe}(\vec{x},t)$. Therefore the vacuum current is gauge invariant in the functional Schrödinger approach. Note that this result is in direct contrast to the canonical approach, where is was shown that, at the formal level, that the canonical formulation of the theory is not gauge invariant.

**5. Conclusion**

We have examined two different formulations of a simple field theory and examined each one to see if it is gauge invariant. The canonical approach, which was introduced in Section 2, was shown not to be gauge invariant at the formal level. This result is consistent with the fact that non-gauge invariant terms occur in the polarization tensor. The functional Schrödinger approach was then examined. It was shown that the vacuum current is gauge invariant if this approach is used. It is interesting to note that the different approaches yield different results. This suggests that it may be useful to formulate quantum field theory using the functional Schrödinger approach instead of the canonical approach.

**Appendix**

Prove that $G(t)^2 = I$. First use (4.28) to obtain,

$$G(t)^2 = \left(2Q(t)\left(Q(t)+P_+\right)^{-1} - I\right)\left(2Q(t)\left(Q(t)+P_+\right)^{-1} - I\right) \tag{A.1}$$

This yields,

$$G(t)^2 = 4Q(t)\left(Q(t)+P_+\right)^{-1}Q(t)\left(Q(t)+P_+\right)^{-1} - 4Q(t)\left(Q(t)+P_+\right)^{-1} + I \tag{A.2}$$

Next use $Q(t) = \left(Q(t)+P_+ - P_+\right)$ in the above to obtain,

$$G(t)^2 = -4Q(t)\left(Q(t)+P_+\right)^{-1}P_+\left(Q(t)+P_+\right)^{-1} + I \tag{A.3}$$

Next use (4.32) and (4.16) to obtain $\left(Q(t)+P_+\right)P_+ = P_+$. This yields,

$$P_+ = \left(Q(t)+P_+\right)^{-1}P_+ \tag{A.4}$$

Use this in (A.3) to obtain,

$$G(t)^2 = -4Q(t)P_+\left(Q(t)+P_+\right)^{-1} + I \tag{A.5}$$

Next use (4.32) to obtain $G(t)^2 = I$. This completes the proof.